\documentclass[
    10pt,
 a4paper,
 superscriptaddress,
 twocolumn,
 footinbib,
 showpacs,
 amsmath,
 amsfonts,
 amssymb,
 aps,
 prb,
floatfix,
]{revtex4-2}

\usepackage[utf8]{inputenc}
\usepackage[T1]{fontenc}

\usepackage{newtxtext,newtxmath}

\usepackage[english]{babel}

\usepackage{jabbrv}

\usepackage{amsmath}
\usepackage{mathtools}
\usepackage{amsfonts}
\usepackage{graphicx}   
\usepackage{dcolumn}    
\usepackage{bm}         
\usepackage{physics}    
\usepackage{hyperref}   

\usepackage[capitalize]{cleveref}

\hypersetup{
colorlinks=true,
citecolor=blue,
linkcolor=red,
urlcolor=black
}

\newcommand{\normord}[1]{\vcentcolon\mathrel{#1}\vcentcolon}
\providecommand{\vcentcolon}{\mathrel{\mathop{:}}}

\newcommand{\DD}[1]{\mathrm{D}{#1}}

\newcommand{\e}[0]{\mathrm{e}}
\renewcommand{\i}[0]{\mathrm{i}}

\begin{document}

\title{Entanglement spectrum and quantum phase diagram of the long-range XXZ chain}

\author{J.~T.~Schneider}
\email{jan.schneider@polytechnique.edu}
\affiliation{CPHT, CNRS, Ecole Polytechnique, IP Paris, F-91128 Palaiseau, France}

\author{S.~J.~Thomson}
\altaffiliation[Current address: ]{Dahlem Centre for Complex Quantum Systems, Freie Universität, 14195 Berlin, Germany}
\affiliation{CPHT, CNRS, Ecole Polytechnique, IP Paris, F-91128 Palaiseau, France}

\author{L.~Sanchez-Palencia}
\email{laurent.sanchez-palencia@polytechnique.edu}
\affiliation{CPHT, CNRS, Ecole Polytechnique, IP Paris, F-91128 Palaiseau, France}

\date{\today}

\begin{abstract}
    Entanglement is a central feature of many-body quantum systems and plays a unique role in quantum phase transitions.
    In many cases, the entanglement spectrum, which represents the spectrum of the density matrix of a bipartite system, contains valuable information beyond the sole entanglement entropy.
    Here we investigate the entanglement spectrum of the long-range XXZ model. We show that within the critical phase  it exhibits a remarkable self-similarity.
    The breakdown of self-similarity and the transition away from a Luttinger liquid is consistent with renormalization group theory.
    Combining the two, we are able to determine the quantum phase diagram of the model and locate the corresponding phase transitions. Our results are confirmed by numerically-exact calculations using tensor-network techniques.
    Moreover, we show that the self-similar rescaling extends to the geometrical entanglement as well as the Luttinger parameter in the critical phase.
    Our results pave the way to further studies of entanglement properties in long-range quantum models.
\end{abstract}

\maketitle

\section{\label{sec-intro}Introduction}

The physical properties of macroscopic systems are determined by the complex interplay of frustrating microscopic interactions and competing symmetries, leading to a variety of critical phenomena.
The traditional approach to phase transitions consists of identifying singular behaviors in local order parameters and two-point correlation functions when a microscopic parameter is continuously varied across the critical point~\cite{sachdev_quantum_2011}.
Such an approach is, however, inoperant for certain quantum phase transitions, including infinite-order and topological transitions, which are signalled only in global quantities~\cite{dechiaraGenuineQuantumCorrelations2018}.
The latter include nonlocal winding numbers and are readily accessible to advanced numerical many-body approaches,
such as quantum Monte Carlo or density-matrix renormalization group methods, see for instance
Refs.~\cite{ceperley1995,schollwock_density-matrix_2011}
and references therein.
It has been recently understood that entanglement properties constitute a fruitful alternative to the characterization of quantum phases and quantum phase transitions~\cite{osterlohScalingEntanglementClose2002, osborneEntanglementSimpleQuantum2002}.
For instance, the von Neumann entropy has been shown to display characteristic logarithmic divergence at critical points~\cite{dechiaraGenuineQuantumCorrelations2018, vidalEntanglementQuantumCritical2003,hastingsAreaLawOnedimensional2007, calabreseEntanglementEntropyQuantum2004,holzheyGeometricRenormalizedEntropy1994, korepinUniversalityEntropyScaling2004}.
Moreover, other entanglement witnesses, such as the geometric entanglement~\cite{barnumMonotonesInvariantsMultiparticle2001, shimonyDegreeEntanglement1995,weiGeometricMeasureEntanglement2003, weiGlobalEntanglementQuantum2005,orusUniversalGeometricEntanglement2008} have been shown to be instrumental for the detection of elusive quantum phase transitions~\cite{orus_visualizing_2010, shiFinitesizeGeometricEntanglement2010, stephanGeometricEntanglementAffleckLudwig2010}.

The complete set of Schmidt weights associated to the ground-state wave function in a bipartition, known as the entanglement spectrum, contains a wealth of information beyond traditional entanglement witnesses~\cite{liEntanglementSpectrumGeneralization2008,regnault_entanglement_2015,laflorencieQuantumEntanglementCondensed2016} and proves instrumental for detecting quantum phase transitions~\cite{albaBoundaryLocalityPerturbativeStructure2012,calabreseEntanglementSpectrumOnedimensional2008, ciracEntanglementSpectrumBoundary2011,dechiaraEntanglementSpectrumCritical2012, lauchliOperatorContentRealspace2013} and topological order~\cite{liEntanglementSpectrumGeneralization2008, regnault_entanglement_2015,pollmannEntanglementSpectrumTopological2010, fidkowskiEntanglementSpectrumTopological2010}.
In spin models for instance, quantum phase transitions have been signaled by a singular behavior of the Schmidt gap~\cite{dechiaraEntanglementSpectrumCritical2012,leporiScalingEntanglementSpectrum2013} and by degeneracy lifts of higher entanglement spectral lines~\cite{plat_entanglement_2020}.
Entanglement properties of prototypical (short-range) spin models have been extensively studied in connection with many-body physics~\cite{peschelEntanglementEntropyXY2004,franchiniRenyiEntropyXY2007,franchiniEllipsesConstantEntropy2007,ercolessi2012,grayManybodyLocalizationTransition2018,dechiaraGenuineQuantumCorrelations2018}.

In this paper, we study the entanglement spectrum of the long-range, spin-\(1/2\) XXZ chain and show that it contains sufficient information to determine the phase diagram as a function of the anisotropy parameter and the interaction range.
The antiferromagnetic-to-XY and XY-to-ferromagnetic phase transitions are, respectively, characterized by degeneracy lifts and the divergence of the Schmidt gap, similarly to the short-range XXZ model.
Analysis of the entanglement spectrum is also instrumental in identifying a remarkable self-similar property of the XY phase. Its breakdown
signals the onset of genuine long-range effects and the spontaneous breaking of a continuous symmetry, consistent with renormalization group theory.
Our results are confirmed by numerical calculations using tensor-network techniques.
Moreover, we show that the self-similarity observed in the entanglement spectrum extends to other quantities, including the geometrical entanglement and the Luttinger parameter in the critical phase.

The paper is organized as follows.
In \cref{sec-model}, we introduce the model and lay out our approach. \cref{sec-geo-ent} studies the quantum phase diagram inferred from the geometric entanglement.
In \cref{sec-ES}, we discuss entanglement-spectrum signatures of quantum phase transitions in the same phase diagram, and we reveal its self-similarity upon rescaling the anisotropic coupling parameter. 
\cref{sec-RG} studies the renormalization group flow of the long-range interacting model in the critical Luttinger phase in combination with the self-similarity thereby confirming the phase diagram.
In \cref{sec-LL}, we verify Luttinger liquid behavior, in particular the self-similarity feature for the Luttinger parameter.
Finally, we draw our conclusion and give an outlook in \cref{sec-conclusion}.

\section{Model and approach\label{sec-model}}

We study the long-range, anisotropic XXZ Heisenberg (LRXXZ) chain,
governed by the Hamiltonian
\begin{align}\label{eq:H-LRXXZ}
 \hat{H} &= -J \sum_{R \neq R'} \frac{ \hat{S}^x_R \hat{S}^x_{R'} + \hat{S}^y_R \hat{S}^y_{R'} + \Delta \hat{S}^z_R \hat{S}^z_{R'}}{\left|R-R'\right|^{\alpha}} \,, 
\end{align}
where \(\hat{S}_R^{j}\) (\(j=x,y,z\)) are the spin-{1}/{2} operators on lattice site \(R \in [0,N-1]\), \(N\) is the system size, \(J>0\) is the coupling energy, and \(\Delta\) is the anisotropy parameter.
As an archetype of an interacting spin chain, the short-range XXZ model has been extensively studied in various contexts such as a textbook example for theoretical techniques like bosonization, the Bethe ansatz~\cite{giamarchi_quantum_2003,takahashi_minoru_thermodynamics_1999,sachdev_quantum_2011,franchini_introduction_2017}, many-body localization~\cite{znidaric_many-body_2008}, as well as out-of-equilibrium dynamics in the context of integrable systems~\cite{essler_quench_2016,calabrese_introduction_2016}.
The short-range anisotropic XXZ chain (\(\alpha \rightarrow \infty\), \textit{i.e.}\ \(\alpha^{-1}=0\)) is integrable
and can be exactly solved via Bethe ansatz~\cite{takahashi_minoru_thermodynamics_1999,sachdev_quantum_2011,franchini_introduction_2017}.
At equilibrium, it encompasses three phases: One finds
a trivial, fully polarized, gapped ferromagnetic (FM) phase for \(\Delta > 1\),
a gapless paramagnetic XY phase for \( -1 < \Delta < 1\),
and a gapped antiferromagnetic (AFM) phase for \(\Delta < -1\).
While the phase transition from XY to FM is of first order,
that from AFM to XY is conversely an 
infinite-order phase transition of the Berezinskii-Kosterlitz-Thouless (BKT) type and no local correlation measure signals this phase transition~\cite{takahashi_minoru_thermodynamics_1999,dechiaraGenuineQuantumCorrelations2018} highlighting the need for a global measure.
Furthermore, an effective low energy description in terms of a conformal field theory (CFT) in the form of a Luttinger liquid can be applied in the paramagnetic phase when interactions are short-ranged~\cite{giamarchi_quantum_2003,sachdev_quantum_2011}.
Although long-range interactions break integrability, the Luttinger liquid description is still valid when including long-range interactions as long as one is interested in the low energy behavior and there exists a well-defined thermodynamic limit~\cite{giamarchi_quantum_2003}.
This implies an effective short-range description of the ground state properties in the gapless phase of the LRXXZ.
The interplay of long-range interactions and continuous symmetry breaking (CSB) has been previously studied in the framework of conformal field theory while employing the renormalization group (RG)~\cite{maghrebi_continuous_2017,inoue_conformal_2006}.
Hereafter we study the entanglement properties of the LRXXZ model~(\ref{eq:H-LRXXZ}) for \(\alpha^{-1} >0\) via density matrix renormalization group (DMRG) simulations~\cite{schollwock_density-matrix_2011}.
We use the matrix product state (MPS) formulation and,
unless otherwise stated, the DMRG calculations are performed using open boundary conditions with maximal bond dimension \(\chi_\mathrm{max} = 250\).

\section{\label{sec-geo-ent} Geometric entanglement}
\begin{figure}
    \includegraphics[width=\linewidth]{{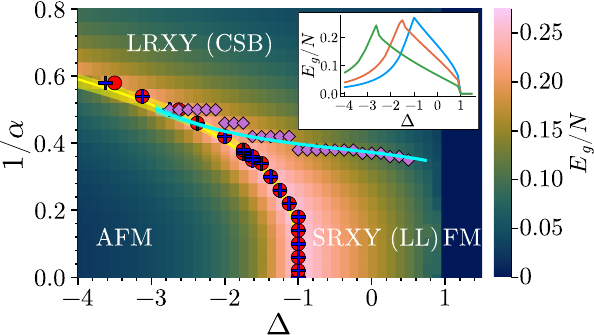}}
    \caption{\label{fig:geo-ent-phase-diag}
        Quantum phase diagram of the LRXXZ model, \cref{eq:H-LRXXZ}, vs the anisotropy (\(\Delta\)) and long-range (\(\alpha^{-1}\)) parameters.
        The color scale indicates the GE density \(E_g/N\), \cref{eq:geo_ent}.
        It shows a cusp at the AFM-XY phase transition (indicated by the red points) and a non-analytic step at the XY-FM phase transition (see Inset). 
        Also shown are numerical results for the AFM-XY phase transition found from degeneracy lift of the entanglement spectrum (blue crosses), and for the upper bound in \(\alpha^{-1}\) for LL behavior (purple diamonds).
        The yellow solid line shows the AFM-XY phase transition as found from renormalization group analysis combined with inverse rescaling of \cref{eq:rescaling} at the critical Luttinger parameter \(K_\mathrm{c}=1/2\), see Sec.~\ref{sec-RG}.    
        The cyan solid line shows the critical line for breaking of LL behavior obtained similarly at \(K'_\mathrm{c}=1/[2(3-\alpha)]\).
        Inset: GE vs \(\Delta\) for \(\alpha^{-1} = 0\) (blue), \(0.34\) (orange), and \(0.5\) (green).
        For all calculations, the system size is \(N=192\).
        The transparent yellow and cyan ribbons as well as the size of the markers correspond to the uncertainty.     
    }
\end{figure}
We first consider the ground-state geometric entanglement (GE),
defined as,
\begin{align}\label{eq:geo_ent}
 E_{g}(\psi) &= - \log_2\left(\max_{\phi : {\mathrm{prod}}} \abs{\braket{\phi}{\psi}}^2\right) \,,
\end{align}
where \(\ket{\psi}\) is the exact ground state of the model and the set \(\{\ket{\phi}\}\) span the submanifold of product states~\cite{weiGeometricMeasureEntanglement2003}.
The GE measures the geometrical distance in Hilbert space of a state to the closest product state.
It has been previously shown to be instrumental for identifying quantum phase transitions, including infinite order ones, in a variety of models, for instance short-range spin models~\cite{orus_visualizing_2010}, and two-dimensional classical models~\cite{gooldTotalCorrelationsDiagonal2015}, and is related to other geometric measures of the entanglement~\cite{schmollClassicalTwodimensionalHeisenberg2021}. 
Here we compute the GE for the LRXXZ model on the MPS ground state using a two-step approach.
We first compute the exact MPS ground state \(\ket{\psi}\) using DMRG calculations with high bond dimension,
perform a singular value decomposition (SVD),
and truncate it down to a single dominating singular value, to reduce the MPS to a product state \(\ket{\phi_0}\).
We then submit the obtained state \(\ket{\phi_0}\) to several variational optimization sweeps, keeping the bond dimension fixed at unity, until convergence of the overlap \(\abs{\braket{\phi}{\psi}}^2\)~\footnote{More precisely, the optimization procedure is stopped when the difference of the overlaps \(\abs{\braket{\phi}{\psi}}^2\) before and after the optimization sweeps becomes negligible (in practice smaller than \(10^{-9}\))}.
This algorithm yields the state \(\ket{\phi}\) closest to the exact ground state \(\ket{\psi}\) within the product-state manifold.

The GE of the LRXXZ model is represented in color scale vs the anisotropy parameter \(\Delta\) and the long-range parameter \(\alpha \)
in \cref{fig:geo-ent-phase-diag}.
The XY to FM phase transition is signaled by a sharp step from \(E_g(\Delta>1) = 0\) to \(E_g(\Delta \leq 1) > 0\),
see inset of \cref{fig:geo-ent-phase-diag}. This transition occurs at \(\Delta=1\), irrespective to the value of the long-range parameter \(\alpha^{-1}\). 
This is consistent with the expected transition to the trivial, fully \(z\)-polarized ground state of the FM phase, see \cref{app:fm-groun-state}.
On the other hand, the AFM to XY phase transition is signaled with a marked cusp of the GE,
see inset of \cref{fig:geo-ent-phase-diag}.
This was previously shown in the short-range XXZ model~\cite{orus_visualizing_2010}, and we find that this feature remains in the long-range case for all considered values of \(\alpha^{-1}\).
We have checked that the position of the local maximum signaling the phase transition is well converged in system size, see \cref{app:finite-size}.
It allows us to locate the XY-FM phase transition vs the long-range parameter, see red disks in \cref{fig:geo-ent-phase-diag}.
While long-range interactions do not affect the XY-FM transition, they significantly shift the AFM-XY transition towards higher values of the antiferromagnetic anisotropy parameter \(\vert\Delta\vert\). This is to be expected since \(z\)-oriented AFM order is frustrated by long-range interactions, which hence favor the XY phase.
The phase diagram as obtained from the GE is in excellent quantitative agreement with that found using the central charge in the conformally symmetric XY phase~\cite{maghrebi_continuous_2017}.
This shows that the GE provides a robust probe of both quantum phase transitions also in the long-range case.

\section{\label{sec-ES}Entanglement spectrum}
To gain more insight into the entanglement properties of the LRXXZ model,
we now study the entanglement spectrum (ES)~\cite{liEntanglementSpectrumGeneralization2008}.
Its properties have been shown to signal quantum phase transitions in a variety of models, including the infinite order BKT phase transition of the XXZ model in the short-range case~\cite{plat_entanglement_2020}.
This contrasts with standard entanglement witnesses, such as Rényi entropies, which show a smooth behavior at the AFM-XY transition~\cite{dechiaraGenuineQuantumCorrelations2018,orus_visualizing_2010}.
The ES is defined from the Schmidt decomposition of
the ground state \(\ket{\psi}\),
\begin{equation}\label{eq:schmidt-decomp}
\ket{\psi} = \sum_j \sqrt{\lambda_j} \ket*{\psi_j^A} \otimes \ket*{\psi_j^B} \,,
\end{equation}
where \(\lambda_j\) is the \(j\)-th Schmidt coefficient, and
\(\ket*{\psi_j^A}\) and \(\ket*{\psi_j^B}\) span an orthonormal basis of each subsystem.
The reduced density matrix of a partition,
\(\rho_A=\tr_B(\ketbra*{\psi})\), is then cast in thermal-like form,
\begin{equation}\label{eq:density_matrix}
\rho_A = \sum_j \e^{-\xi_j} \ketbra*{\psi_j^A},
\end{equation}
where the coefficients \(\xi_j = -\ln(\lambda_j)\) are the entanglement energies and the effective temperature equals unity.

\begin{figure}
    \centering
    \includegraphics[width=0.9\linewidth]{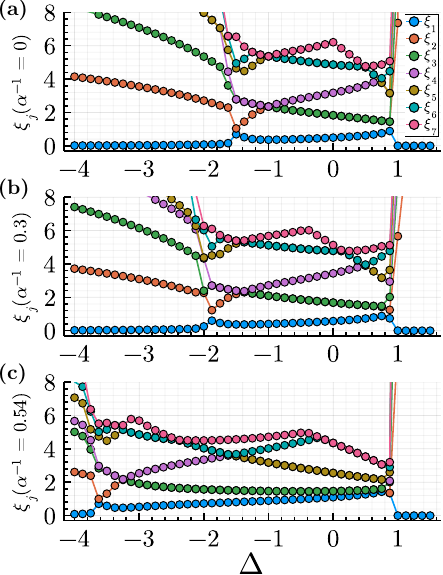}
    \caption{\label{fig:LRXXZEE_bare}
        Entanglement spectrum of the ground state of
        (a)~the short-range XXZ model, \(\alpha^{-1}=0\),
        and (b)-(c) the LRXXZ model for \(\alpha^{-1}=0.3\) and \(\alpha^{-1}=0.54\), respectively.
        Shown are the first seven entanglement energies (\(\xi_1\) blue, \(\xi_2\) orange, \(\xi_3\) green, \(\xi_4\) purple, \(\xi_5\) brown, \(\xi_6\) teal, and \(\xi_7\) pink).
        The phase transition from XY to FM is marked by the divergence of all but the first entanglement energies at \(\Delta=1\).
        In the XY phase, the entanglement energies \(\xi_2\) (orange) and \(\xi_3\) (green) are degenerate (\(\xi_2=\xi_3\)) while they are distinct deep in the AFM phase (\(\xi_2 \neq \xi_3\)), and the degeneracy lift marks the AFM--XY phase transition.
        Note the ES is in a crossover regime between the AFM--XY critical point and the cusps of \(\xi_1\) and \(\xi_2\), see text.
        The system size is \(N=192\) for all calculations.
        The error bars are smaller than the size of the markers.
    }
\end{figure}

\subsection{\label{sec-ES-results}Entanglement spectrum of the LRXXZ chain}

The ground-state ES of the LRXXZ is shown in Fig.~\ref{fig:LRXXZEE_bare} vs the anisotropy parameter \(\Delta\)
in the short-range case [(a)~\(\alpha^{-1}=0\)]
and in the long-range case for two values of the long-range parameter [(b)~\(\alpha^{-1}=0.3\) and (c)~\(\alpha^{-1}=0.54\)].
In all cases, the XY to FM phase transition is marked by the sharp divergence of all entanglement energies but \(\xi_1\), which vanishes for \(\Delta=1\).
This is consistent with the onset of a fully polarized, exact product state in the FM phase, irrespective of the long-range parameter \(\alpha^{-1}\).
The ground state deep in the AFM phase also tends towards a product state but only smoothly in the limit of infinite anisotropy, \(\Delta \rightarrow -\infty \),
as indicated by the monotonous increase of all \(\xi_j\) but \(\xi_1\).
In the short-range case, Fig.~\ref{fig:LRXXZEE_bare}(a), the AFM to XY phase transition at \(\Delta=-1\) is marked by the
sudden lift in the degeneracy of entanglement energies, see also Ref.~\cite{plat_entanglement_2020}.
More precisely, the entanglement energies \(\xi_2\) and \(\xi_3\) are degenerate in the XY phase
while they are distinct in the AFM phase.  The degeneracy lift, found exactly at  \(\Delta = -1\), marks the AFM--XY phase transition.
A similarly sharp degeneracy lift is found for the  entanglement energies \(\xi_5\) and \(\xi_6\).
Qualitatively similar features are found in the long-range case, for all considered values of long-range parameter \(\alpha^{-1}\). The degeneracy lift point is, however, found for a critical anisotropy parameter \(\Delta\) that significantly depends on the long-range parameter \(\alpha^{-1}\), Figs.~\ref{fig:LRXXZEE_bare}(b) and \ref{fig:LRXXZEE_bare}(c).
It allows us to locate the AFM to XY phase transition in the LRXXZ model for all values of the long-range parameter.
The result, shown as blue crosses in the phase diagram of Fig.~\ref{fig:geo-ent-phase-diag}, is in excellent agreement with the transition previously inferred from the cusp of the GE.

Note that the ES shows an apparent crossover regime in a narrow region of the AFM phase close to the AFM--XY phase transition, even for the relatively large system size used in our calculations (\(N = 192\)). It is marked by apparent degeneracies (e.g.\ \(\xi_3 = \xi_4\)) and a cusp of the lowest two entanglement energies in this crossover regime, see behavior in the interval \( -1.5 < \Delta < -1\) for the short-range case, and lower values for long-range cases.
However, we find that this interval slowly shrinks towards the true AFM--XY critical point, see Appendix~\ref{app:finite-size}.
This is consistent with the slow finite-size scaling of the local maximum of the entanglement entropy reported in earlier papers~\cite{wang_berezinskii-kosterlitz-thouless_2010,li_ground_2019}.
For a more detailed discussion of the finite-size effects on the ES, see \cref{app:finite-size}.
We hence consider the cusps of \(\xi_1\) and \(\xi_2\), i.e., the local minimum of the Schmidt gap, as well as the apparent degeneracies in the crossover regime observed in Fig.~\ref{fig:LRXXZEE_bare} as finite-size artifacts.
Note that in striking contrast the degeneracy lift point is nearly independent of the system size for \(N \gtrsim 100\), see Appendix~\ref{app:finite-size}.

\subsection{\label{sec-ES-self-similarity}Self-similarity}

\begin{figure}[tp]
    \centering
    \includegraphics[width=0.9\linewidth]{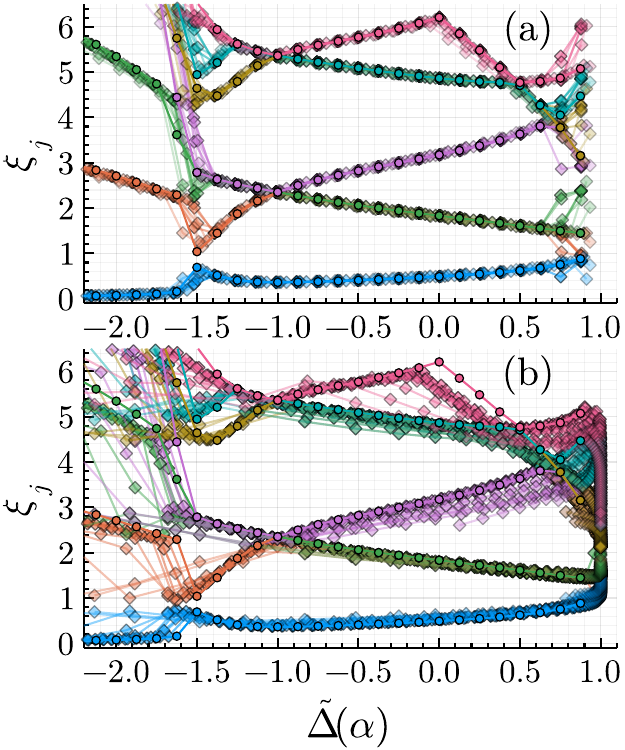}
    \caption{\label{fig:EE-rescaled}
        Entanglement spectra plotted vs the rescaled anisotropy parameter  \(\tilde{\Delta}\) for various long-range parameters \(\alpha^{-1}\).
        The short-range case \(\alpha^{-1} = 0\) is shown as colored circles while long-range cases are shown as colored {diamonds}.
        (a):~Values of \(\alpha^{-1}\) from \(0.02\) to \(0.3\) with an increment of \(0.04\) corresponding to progressively fainter color.
        (b):~Same for values of \(\alpha^{-1}\) from \(0.34\) to \(0.78\).
    }
\end{figure}

\begin{figure}
    \centering
    \includegraphics[width=0.9\linewidth]{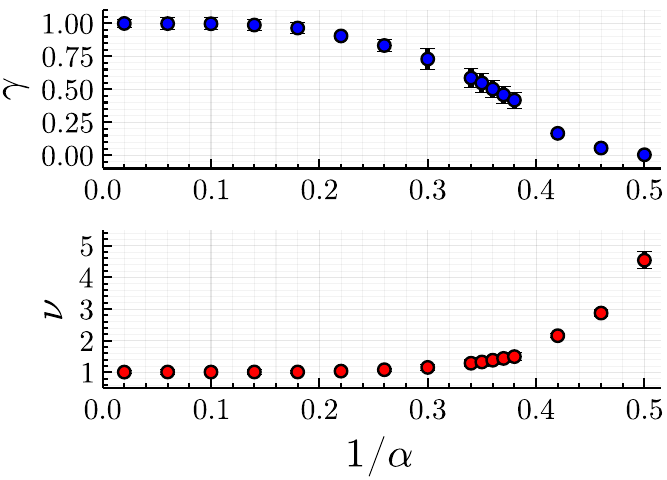}
    \caption{
        Rescaling parameters \(\gamma\) and \(\nu\) of \cref{eq:rescaling} vs the long-range parameter \(\alpha^{-1}\).
        The error bars are indicated by capped black lines (see method in \cref{app:optim-parameters}).
    }
    \label{fig:rescaling-params}
\end{figure}

Inspection of the ES for a variable interaction range in the various panels of \cref{fig:geo-ent-phase-diag} shows a remarkable similarity, in particular in the low-entanglement energy sector of the XY phase.
More precisely, we can find a nonlinear rescaling of the anisotropy parameter of the form
\begin{equation}\label{eq:rescaling}
    \Delta \rightarrow \tilde{\Delta}(\Delta,\alpha) = -\gamma(\alpha) \abs{ \Delta - 1}^{\nu(\alpha)} +1 \,,
\end{equation}
such that all spectral lines (approximately) collapse onto the ES of the short-range model, see \cref{fig:EE-rescaled}.
The rescaling~(\ref{eq:rescaling}) is consistent with the exact fixed point \(\Delta^\star = 1\) corresponding to the XY--FM transition.
The parameters \(\gamma(\alpha)\) and \(\nu(\alpha)\) are then found by
minimizing a weight function constructed over a wide interval containing the AFM--XY phase transition.
For more details, see \cref{app:optim-parameters}.
\Cref{fig:rescaling-params} shows the rescaling parameters \(\gamma\) and \(\nu\) vs the long-range parameter \(\alpha^{-1}\).
Both start to significantly differ from unity at \(\alpha^{-1} \simeq 0.2\), consistently with Fig.~\ref{fig:geo-ent-phase-diag}.

More precisely, we distinguish two regimes.
For roughly \(\alpha^{-1} \lesssim 0.3\), all spectral lines almost perfectly collapse onto the short-range ES upon rescaling in the XY phase, see Fig.~\ref{fig:EE-rescaled}(a).
On the other hand, for \(\alpha^{-1} \gtrsim 0.3\), only the lowest three spectral lines \(\xi_1,\xi_2,\xi_3\) are congruent with the short-range ones upon rescaling, see Fig.~\ref{fig:EE-rescaled}(b).
In contrast, the rescaled spectrum shows a worse match for higher entanglement energies (\(\xi_4\) and higher).
For instance, while \(\xi_4\) (purple) is rescaled to greater values of \(\tilde{\Delta}\), \(\xi_7\) (pink) is rescaled to lower ones, pointing towards an irreconcilable mismatch following a global rescaling of the ES.
Nevertheless, the good match of the lowest entanglement energies renders the scaling~(\ref{eq:rescaling}) sufficient to determine the AFM to XY transition found from the degeneracy lift of \(\xi_2\) and \(\xi_3\).
Similarly, the rescaling parameters \(\gamma(\alpha)\), \(\nu(\alpha)\) start to significantly deviate from close to unity around \(\alpha^{-1} \simeq 0.2\) reflecting an onset of a stronger expansion of the ES to more negative values of \(\Delta\) with longer ranged interactions. See also \cref{fig:LRXXZEE_bare} for a direct observation of the degeneracy lift position expanding faster than linear towards more negative \(\Delta\) with longer ranged interactions.

\begin{figure}[htp]
    \centering
    \includegraphics[width=0.9\linewidth]{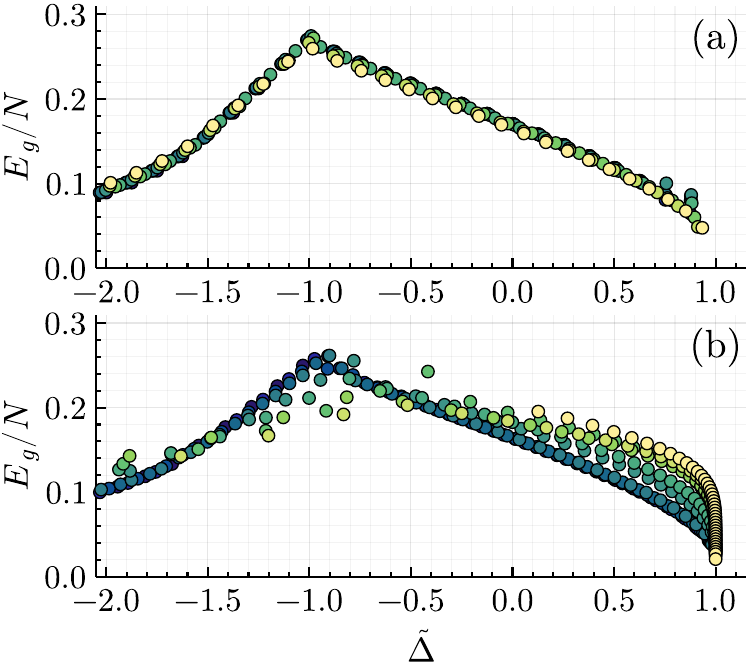}
    \caption{
     Rescaling of the ground-state GE density, \cref{eq:geo_ent}, with \cref{eq:rescaling} vs \(\tilde{\Delta}\).
    (a)~Data for \(\alpha^{-1} \leq 0.3\), color coding from dark blue to bright yellow for \(\alpha^{-1} = [0.0, 0.02, 0.06, \ldots, 0.3]\).
    (b)~Data for \(\alpha^{-1} > 0.3\), color coding from dark blue to bright yellow for \(\alpha^{-1} = [0.34, 0.38, \ldots, 0.62]\).
    }\label{fig:geo-ent-self-similar}
\end{figure}

Likewise, we have checked that the same rescaling also applies to the geometric entanglement curves.
Applying the rescaling of \cref{eq:rescaling} with the scaling parameters \(\gamma(\alpha)\) and \(\nu(\alpha)\) found from the ES (\cref{fig:rescaling-params}), we find very good data collapse of the rescaled GE curves onto the corresponding short-range curve for \(\alpha^{-1} \lesssim 0.3\), see~\cref{fig:geo-ent-self-similar}(a). It is worth noting that this holds over both the AFM and XY phase.
In contrast, for \(\alpha^{-1} \gtrsim 0.3\), the rescaling gets increasingly worse for longer range interactions (increasing values of \(\alpha^{-1}\)), although the cusp is still consistent with \(\tilde{\Delta} \simeq -1\), see~\cref{fig:geo-ent-self-similar}(b).

\section{\label{sec-RG}Renormalization group analysis of quantum phase diagram}

The breakdown of ES self-similarity around \(\alpha^{-1}\sim 0.3\) suggests a transition towards a phase belonging to a universality class different from the short-range XY phase. This is consistent with the continuous symmetry breaking phase transition identified in Ref.~\cite{maghrebi_continuous_2017}.
There, the transition was found using DMRG calculations and perturbative renormalization group (RG) analysis was shown to fairly predict the transition around the XY point (\(\Delta \simeq 0\)) in spite of significant deviations from the numerical results.

Below we show that combining RG theory with the anisotropic parameter rescaling allows us to precisely identify the transition in excellent agreement with numerical calculations.
As a starting point, we apply standard perturbative RG theory, working along the lines of Refs.~\cite{giamarchi_quantum_2003,sachdev_quantum_2011,franchini_introduction_2017,dutta_phase_2001,defenu_criticality_2017,maghrebi_continuous_2017}.
Within the XY phase, the  short-range XXZ model is well described by Luttinger liquid (LL) theory.

The latter is characterized by a massless, quadratic, conformally-invariant field theory with central charge \(c=1\), described by the Hamiltonian
\begin{equation} \label{eq:H-LL}
    H_\mathrm{LL} = \frac{u}{2} \int \dd{x} \left[ K \big( \partial_x \theta(x) \big)^2 + \frac{1}{K} \big( \partial_x \phi(x)\big)^2 \right]\,,
\end{equation}
where \(\phi(x)\) is a scalar field, \(\Pi(x) = \partial_x \theta(x)\) is the canonical conjugate momentum with \(\comm*{\Pi(x)}{\phi(y)} = \mathrm{i} \delta(x-y)\), \(\theta(x)\) is the dual field to \(\phi(x)\),
\(K\) is the Luttinger parameter, and \(u\) is the speed of sound.
We then include the field-theoretic interaction terms corresponding to the microscopic 
short-range \(S^z_n S^z_{n+1}\) (ZZ) interaction,
as well as the long-range \(S^x_n S^x_m + S^y_n S^y_m\) (LRXY) and \(S^z_n S^z_m\) (LRZZ) interactions.
Ignoring oscillating factors, they yield three terms:
\begin{align}
    V_\mathrm{ZZ}[\phi] &= - \frac{g_\mathrm{ZZ}}{(\pi a_{\mathrm{c}})^2} \int  \cos(\sqrt{16\pi} \phi(x)) \dd{x} \,, \label{eq:V-ZZ}\\
    V_\mathrm{LRXY}[\theta] &=  - \frac{g_\mathrm{LR}}{2\pi a_{\mathrm{c}}} \int' \frac{\cos(\sqrt{{\pi}} (\theta(x)-\theta(y)) )}{\abs*{x-y}^\alpha}  \dd{x} \dd{y} \,, \label{eq:V-LRXY} \\
    V_\mathrm{LRZZ}[\phi] &=  - \frac{g_\mathrm{LRZZ}}{\pi} \int' \frac{\partial_x \phi(x) \partial_y \phi(y)}{\abs*{x-y}^\alpha}  \dd{x} \dd{y} \,, \label{eq:V-LRZZ}
\end{align}
where the integral \(\int' \dd{x}\dd{y}\) runs over \(\abs*{x-y}\gg a_c\) and \(a_{\mathrm{c}}\) is the ultraviolet cutoff until the LL theory approximation holds on the spin lattice.
The long-range ZZ interaction term \(V_\mathrm{ LRZZ}[\phi]\), Eq.~\eqref{eq:V-LRZZ}, is omitted here
since it is irrelevant for \(\alpha^{-1} <1\)~\cite{inoue_conformal_2006}.
The microscopic action is thus given by \(S[\phi,\theta] = S_0[\phi] -V_\mathrm{ZZ}[\phi] - V_\mathrm{LRXY}[\theta]\), with the free quadratic part of the action \(S_0[\phi] = {(2 K)}^{-1} \int \dd[2]{\mathbf{r}} \left( \partial_\mu \phi(\mathbf{x}) \right)^2\) where \(\mathbf{r} = r^\mu = (u \cdot t, x)\) is the spacetime vector.
RG theory then yields the flow equations
\begin{align}
    \dv{g_\mathrm{ ZZ}}{\ell} &= \left(2 - 4K\right) g_\mathrm{ ZZ} \,, \label{eq:rg-flow-zz} \\
    \dv{g_\mathrm{ LR}}{\ell} &= \left(3 - \alpha - \frac{1}{2K}\right) g_\mathrm{ LR} \,, \label{eq:rg-flow-lr}
\end{align}
where \(\dd{\ell}\) is the width of the spacetime momentum shell at momentum cutoff \(\Lambda\), integrated out in one RG step. For more details and a derivation, see \cref{app:RG-flow}.
Note that the flow equations (\ref{eq:rg-flow-zz}) and (\ref{eq:rg-flow-lr}) are uncoupled.
To first-order perturbative RG, the renormalization of \(g_\mathrm{ ZZ}\) is thus completely controlled by the Luttinger parameter \(K\)~\cite{giamarchi_quantum_2003,shankar_quantum_2017},
while that of \(g_{\mathrm{ LR}}\) is controlled by both \(K\) and the long-range exponent \(\alpha\)~\cite{maghrebi_continuous_2017}.

Equation~\eqref{eq:rg-flow-zz} governs the AFM-XY transition:
The ZZ term is relevant for \(K < 1/2\) (AFM phase) and irrelevant for \(K >1/2\) (XY phase), indicating a transition characterized by the critical Luttinger parameter \(K_\mathrm{c}=1/2\).
This prediction is confirmed by numerical calculations all along the AFM-XY transition line, see Sec.~\ref{sec-LL}.
In the short-range case, the corresponding critical anisotropy parameter \(\Delta_c\) is readily found relying on the Bethe ansatz formula, 
\begin{equation}\label{eq:BAK}
K(\Delta, \alpha=\infty) = {\pi}/[{2\arccos(\Delta)}],
\end{equation}
valid in the XY phase,
which yields
\(\Delta_{\mathrm{c}}(\alpha = \infty) = -1\).
In the long-range case, the critical AFM--XY line in the \(\alpha\)-\(\Delta\) plane may then be found by solving \(\tilde{\Delta}(\Delta_\mathrm{c},\alpha)=-1\) for \(\Delta_\mathrm{c}\) in Eq.~(\ref{eq:rescaling}).
This yields the solid yellow line in the phase diagram of Fig.~\ref{fig:geo-ent-phase-diag}, which shows very good agreement with the estimates based on either the GE cusp (red points) or the direct degeneracy lift of the ES
(blue diamonds).

The rescaling of the ES suggests that not only the critical AFM-XY point
can be rescaled to its long-range counterpart, but rather can \(K\) be rescaled over a wider region of the XY phase.
With this assumption, we inspect the critical line where \(V_\mathrm{ LRXY}[\theta]\), Eq.~\eqref{eq:V-LRXY}, turns relevant, corresponding to the condition \(K'_{\mathrm{c}} = 1/[2(3-\alpha)]\), see Eq.~(\ref{eq:rg-flow-lr}). Replacing \(K'_{\mathrm{c}}\) by the Bethe ansatz formula~(\ref{eq:BAK}) and \(\Delta\) by \(\tilde{\Delta}(\Delta'_{\mathrm{c}},\alpha)\), Eq.~(\ref{eq:rescaling}), we then solve for \(\Delta'_{\mathrm{c}}(\alpha)\).
It yields the cyan solid line in \cref{fig:geo-ent-phase-diag}. Above this line, the physics is governed by the long-range XY model (LRXY phase in the Fig.~\ref{fig:geo-ent-phase-diag}), while below the long-range XY term is irrelevant (SRXY phase).
Numerical calculations confirm that the transition is characterized by the critical Luttinger parameter \(K'_{\mathrm{c}} = 1/[2(3-\alpha)]\), see Sec.~\ref{sec-LL}.
Note that for \(\alpha=2\), the critical line for long-range behavior yields \(K'_{\mathrm{c}} = 1/2\), at which point also \(V_\mathrm{ ZZ}[\phi]\) becomes relevant (\(K_{\mathrm{c}} = 1/2\)). We thus expect that the two critical lines approximately meet around \(\alpha_{\mathrm{c}}=2\) within first order perturbative RG.

\section{\label{sec-LL}Luttinger liquid parameters in XY phase}

The results above indicate that the entanglement properties (ES and GE) of the LRXXZ model can be deduced from their short-range counterpart upon the rescaling of Eq.~(\ref{eq:rescaling}), at least for moderate long-range interactions (roughly \(\alpha^{-1} \lesssim 0.3\)).
Furthermore, our RG analysis is consistent with the existence of an effective Luttinger parameter \(K\) fulfilling the same rescaling over the entire SRXY phase.
In this section, we check Luttinger liquid behavior as well as the self-similar features of \(K\) over the critical phase by inspecting a number of universal behaviors characteristic of LL.

We first consider the behavior of the Rényi entropies,
\(\mathcal{S}_n = \ln[\tr(\rho_{A}^n)]/(1-n)\) with Rényi order \(n\in \mathbb{R}^+\).
Measuring Rényi entropies allows us on the one hand to verify that the central charge is close to unity, a necessary condition for LL behavior, and on the other hand to estimate the LL parameter \(K\).
The Rényi entropies of the short-range XXZ model in the critical XY phase may be written as~\cite{calabrese_parity_2010,xavier_renyi_2011}
\begin{align}\label{eq:renyi-entropy-theory}
 \mathcal{S}_n(N,l) = \mathcal{S}^{\mathrm{CFT}}_n(N,l) + \mathcal{S}^{\mathrm{osc}}_n(N,l),
\end{align}
with \(N\) the system size and a bipartition into two subsystems  \(A\) and \(B\) of respective sizes \(l\) and \(N-l\).
The first term is the conformal field theory (CFT) prediction.
For a finite one-dimensional gapless system of size \(N\) with open boundary conditions,
it reads as
\begin{align}\label{eq:S_CFT}
 \mathcal{S}^{\mathrm{CFT}}_n = \frac{c \left(1 + {n}^{-1}\right)}{12} \ln\left[\frac{4(N+1)}{\pi} \sin(\frac{\pi(2l + 1)}{2(N+1)})\right] + c_1 \,,
\end{align}
where \(c\) is the central charge, and \(c_1\) is a nonuniversal constant~\cite{calabrese_parity_2010,xavier_renyi_2011}.
Note that the open boundary conditions (OBC) alter the chord distance \(D_\mathrm{OBC}(l,N) =  4(N+1)\sin[{\pi(2l + 1)}/(2(N+1))]/\pi\) with respect to periodic boundary conditions (PBC) \(D_\mathrm{PBC}(l,N) = {N}\sin(\pi l /N)/\pi\)~\cite{xavier_renyi_2011,xavier_finite-size_2012}.
The second term accounts for oscillatory corrections to the CFT prediction due to significant antiferromagnetic correlations in the critical XY phase of the XXZ model~\cite{calabreseUniversalCorrectionsScaling2010,xavier_renyi_2011}.
It takes the universal form~\cite{calabrese_parity_2010, calabreseUniversalCorrectionsScaling2010, xavier_renyi_2011, fagotti_universal_2011, xavier_finite-size_2012}
\begin{align}\label{eq:S_osc}
\mathcal{S}^{\mathrm{osc}}_n = \frac{g_n}{N^{p_n}} \sin[(2l+1)k_F'] \abs{\sin(\frac{\pi(2l +1)}{2(N+1)})}^{-p_n}\,, 
\end{align}
where \(g_n\) is a nonuniversal constant, the exponents of the oscillation amplitude, \(p_n\), are related to the Luttinger parameter \(K\) as \(p_n = 2K / n\) for OBC, and \(k_F' = \frac{N}{N+1} k_F + \frac{\pi}{2(N+1)}\) is an effective Fermi momentum, including OBC finite-size corrections with respect to its counterpart in the thermodynamic limit, \(k_F=\pi/2\).

\begin{figure}
    \centering
    \includegraphics[width=0.9\linewidth]{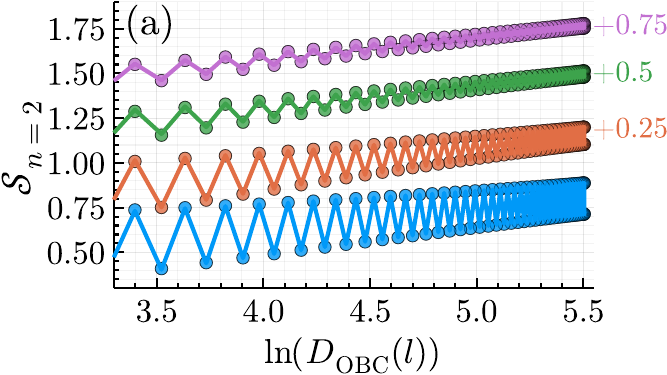} \\
    \includegraphics[width=0.9\linewidth]{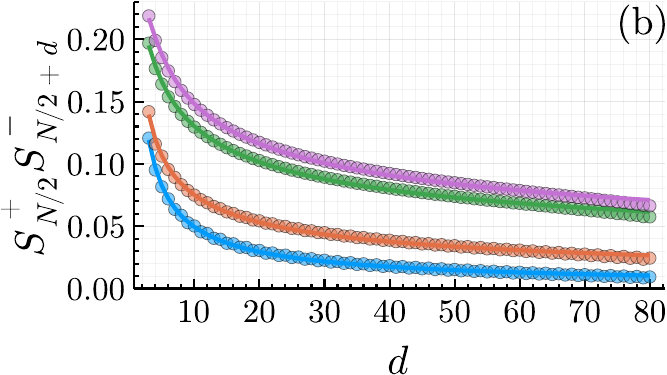}
    \caption{
    Typical MPS data for entanglement entropies and correlation functions for
        \(\Delta=-0.875\), \(\alpha^{-1}=0.06\) (blue),
        \(\Delta=-1.5\), \(\alpha^{-1}=0.42\) (orange),
        \(\Delta=-0.125\), \(\alpha^{-1}=0.34\) (green),
        \(\Delta= 0.5\), \(\alpha^{-1}=0.14\) (purple),
        and the system size \(N=192\).
        The error bars are smaller than the size of the marker.
        (a)~Rényi entanglement entropies \(\mathcal{S}_{n=2}(N,l)\) (colorized circles) and corresponding fit of \cref{eq:renyi-entropy-theory} (solid line) plotted vs the logarithm of the chard length.
        Fits performed over \(l \in [10,\ldots,N-10]\) yield
            \(c_\mathrm{eff} = 0.9\), \(K=0.6\) (blue),
            \(c_\mathrm{eff} = 1\), \(K=1.1\) (orange),
            \(c_\mathrm{eff} = 1\), \(K=1.5\) (green), and
            \(c_\mathrm{eff} = 1\), \(K=1.5\) (purple).
        For clarity, the curves are shifted by a constant offset indicated on the right-hand-side of each curve.
        (b)~Correlation function \(\expval*{S^+_{N/2} S^-_{N/2+d}}\) (points) and corresponding fit of \cref{eq:twopoint-pm} (solid line) plotted vs distance \(d\).
        The fits yield
            \(K=0.6\) (blue),
            \(K=1.1\) (orange),
            \(K=1.4\) (green), and
            \(K=1.4\) (purple).
    }\label{fig:fit_examples}
\end{figure}

To determine the effective central charge \(c\) and Luttinger parameter \(K\) of the LRXXZ, we fit \cref{eq:renyi-entropy-theory} with Eqs.~(\ref{eq:S_CFT}) and (\ref{eq:S_osc}) to the Rényi entropy obtained from the ground state MPS in the range \(l \in [10, \ldots, N-10]\) at fixed system size \(N\) and fixed Rényi order \(n\).
It yields estimates of the four fitting parameters
\(c\), \(c_1\), \(g_n\), \(p_n\),
and consequently of the Luttinger parameter, \(K = n p_n/2\).
We focus on the critical XY phase (\(-1<\tilde{\Delta}<1\)) as previously identified from the GE and ES.
We consider that the Rényi entropies are consistent with LL behavior when the residual sum of squares (RSS) is below \(2\%\).
Typical fits of \cref{eq:renyi-entropy-theory} to the MPS data are displayed in \cref{fig:fit_examples}(a). Note that for clarity the various curves are shifted by an amount indicated on the right-hand-side of each curve.
Judging from the fit quality check above, we find that the results are consistent with LL behavior in a region of the critical phase bounded from above in direction of \(\alpha^{-1}\).
The boundary is displayed as purple diamonds in \cref{fig:geo-ent-phase-diag}.
The breakdown of the LL behavior is in excellent agreement with the critical line found from our RG analysis (solid-cyan line).
Consistently, we find that in the region so identified, the central charge---as extracted from the fits---does not significantly deviate from unity \(c \approx 1\).
This property was used in Ref.~\cite{maghrebi_continuous_2017} as a criterion to identify the LL phase and yields a similar boundary. Moreover, we find that the various estimates of the Luttinger parameter from Rényi entropies of different orders \(n\), \(K=n p_n/2\), consistently yield a value of \(K\) approximately independent of \(n\) with a tolerance of less that \(7.5\%\).

To double-check the validity of the LL behavior within the SRXY phase, we now turn to a second, independent measurement of the effective Luttinger parameter \(K\).
To this end, we consider the \(\expval*{S^+_R S^-_{R'}}\) correlation functions.
For the short-range XXZ model they read as,
\begin{align}\label{eq:twopoint-pm}
G(R,R') = \expval{S^+_R S^-_{R'}} &\approx  \frac{C_1}{\abs{R-R'}^{2K + \frac{1}{2K}}} +  \frac{C_2}{\abs{R-R'}^{\frac{1}{2K}}} \,,
\end{align}
where \(C_1\) and \(C_2\) are nonuniversal constants~\footnote{The Hamiltonian of the short-range model obeys the symmetry \(H(-J,-\Delta,h,\alpha=\infty) = U_2 H(J,\Delta,h,\alpha=\infty) U_2^{-1}\) with \(U_2 = \prod_{l=\mathrm{even}} \sigma^z_l = U_2^{-1}\). This maps \(S^x_l\), \(S^y_l\), \(S^z_l\) to \(-S^x_l\), \(-S^y_l\), \(S^z_l\) for even \(l\)~\cite{takahashi_minoru_thermodynamics_1999}.
Using this transformation, \cref{eq:twopoint-pm} is consistent with its counterpart for antiferromagnetic coupling (\(J<0\)), see Refs.~\cite{takahashi_minoru_thermodynamics_1999,giamarchi_quantum_2003}.}.
We compute the correlation function \(\expval*{S^+_{N/2} S^-_{N/2+d}}\) in the LL regime identified above.
Typical MPS results (points) together with fits of Eq.~(\ref{eq:twopoint-pm}) to the data (solid lines) are shown in \cref{fig:fit_examples}(b), showing excellent agreement over the full LL regime.
These fits confirm the LL behavior and yield a second, independent, estimate of the Luttinger parameter \(K\).

To compare the two estimates of the Luttinger parameter, from fits to the Rényi entropies (\(K^\mathcal{S}\)) and to the correlation functions (\(K^G\)) respectively, first note that the same congruent rescaling as discussed above also applies to both \(K\) estimates.
\begin{figure*}
    \centering
    \includegraphics[width=0.85\linewidth]{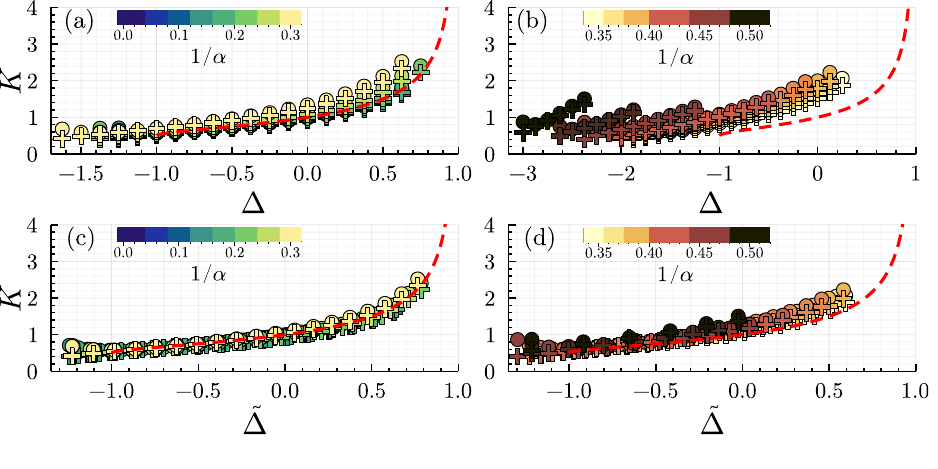}
    \caption{\label{fig:luttinger-K} 
        Luttinger parameter \(K\) vs the anisotropy parameter for various values of the long-range parameter \(\alpha^{-1}\).
        Colored disks correspond to the Rényi-entropy estimates \(K^{\mathcal{S}}\), \cref{eq:renyi-entropy-theory} with Rényi order \(n=2\), and colored crosses to correlation-function estimates \(K^{G}\), \cref{eq:twopoint-pm}.
        Color codings for \(\alpha^{-1}\) are indicated in color bars.
        The dashed red line is the short-range Bethe ansatz formula~(\ref{eq:BAK}).
        The upper panels~[(a), (b)] show the data vs the bare anisotropy parameter \(\Delta\), separating the cases \( 0 \leq \alpha^{-1} \leq 0.3\) and \( \alpha^{-1} >0.3\) for clarity.
        The lower panels~[(c), (d)] show, respectively, the same data vs the rescaled parameter \(\tilde{\Delta}\).
        The breakdown of LL theory is signaled by deviations between the two estimates at \(\tilde{\Delta}\lesssim -1\).
        The size of the error bars is smaller than the size of the marker.
    }
\end{figure*}
Figures~\ref{fig:luttinger-K}(a) and (b) show the fitted values of \(K^\mathcal{S}\) (colored dots) and \(K^G\) (colored stars) plotted against the anisotropy parameter \(\Delta\) for different values of the long-range parameter \(\alpha^{-1}\).
Data for \(0<\alpha^{-1} \leq 0.3\) and \(0.3<\alpha\), respectively, are separated in Figs.~\ref{fig:luttinger-K}(a) and (b) for clarity.
The dashed red line shows the analytic short-range result from Bethe ansatz, Eq.~(\ref{eq:BAK}).
Figures~\ref{fig:luttinger-K}(c) and (d) show, respectively, the same data vs the rescaled anisotropy parameter \(\tilde{\Delta}\).
For \(0 \leq \alpha^{-1} \leq 0.3\), we observe an almost perfect collapse of all long-range values of \(K^\mathcal{S}\) and \(K^G\) onto the short-range curve upon the rescaling~\eqref{eq:rescaling} with parameters as in \cref{fig:rescaling-params}, see \cref{fig:luttinger-K}(c).
In contrast, the data for \(K^\mathcal{S}\) and \(K^G\) for \(\tilde{\Delta} < -1\) do not agree with each other, consistently with the breakdown of LL theory.
Similarly, Fig.~\ref{fig:luttinger-K}(d) displays \(K^\mathcal{S}\) and \(K^G\) for \(0.3 < \alpha^{-1}\) (color coding from blue to yellow) and shows a good self-similar rescaling,
albeit with few singular deviations inside \(-1 < \tilde{\Delta} < 1\) for the largest values of \(\alpha^{-1}\).
Here also the breakdown of LL theory is found for \(\tilde{\Delta} < -1\) where \(K^\mathcal{S}\) and \(K^G\) deviate from each other.

These results confirm the LL behavior in the SRXY phase identified in the diagram of Fig.~\ref{fig:geo-ent-phase-diag}.
They furthermore confirm the self-similarity features of the Luttinger parameter \(K\) that was assumed in the derivation of the critical lines in our RG analysis.
The latter matches well our independent numerical analysis in this section,
as well as the numerical results of Ref.~\cite{maghrebi_continuous_2017}.
The critical line we obtain from RG analysis constitutes a substantial improvement over the perturbatively computed analytic line \emph{ibidem}.

\begin{figure}
    \centering
    \includegraphics[width=\linewidth]{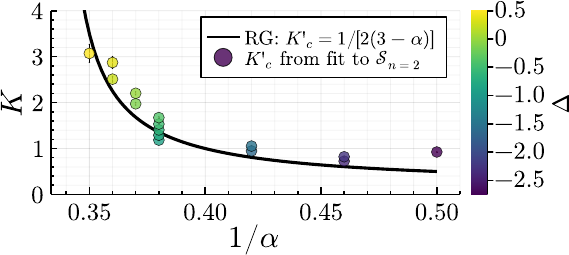}
    \caption{
        Critical Luttinger parameter along the SRXY-LRXY transition vs the long-range parameter \(\alpha^{-1}\).
        The solid black line is the RG prediction \(K'_{\mathrm{c}} = 1/[2(3-\alpha_{\mathrm{c}})]\).
        The colored disks show the numerically measured \(K^{\mathcal{S}}\) through fits to \cref{eq:renyi-entropy-theory} with Rényi order \(n=2\) at points in the phase diagram~\cref{fig:geo-ent-phase-diag} closest to the RG critical boundary line (cyan solid line). The color code correspond to values of \(\Delta\) from \(\Delta = 0.5\) (yellow) to \(\Delta=-2.75\) (blue).
    }
    \label{fig:K-crit-RG}
\end{figure}

Moreover, we find that the point where the two numerically estimated Luttinger parameters start to deviate, hence marking the breakdown of LL behavior and found at \(\tilde{\Delta} \simeq -1\), is consistent with the RG prediction for the AFM--XY phase transition, \(K_c=1/2\), for all values of \(\alpha\) inspected, see~\cref{fig:luttinger-K}(c).
Similarly, we find that the Luttinger parameter computed from Rényi entropies, \(K^\mathcal{S}\), along the phase boundary estimated by \(K_c'\) (solid cyan line in \cref{fig:geo-ent-phase-diag})
is consistent with the RG prediction for the SRXY-LRXY transition, see \cref{fig:K-crit-RG}.

\begin{figure}
    \centering
    \includegraphics[width=\linewidth]{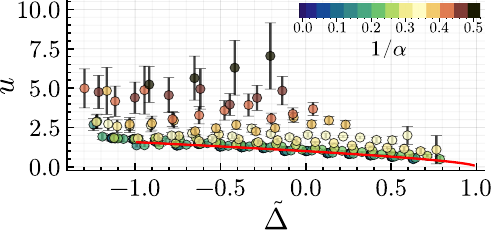}
    \caption{\label{fig:luttinger-u}
        Speed of sound \(u\) vs the rescaled anisotropy parameter \(\tilde{\Delta}\) for various values of the long-range parameter \(\alpha^{-1}\). The data are found from calculations of the magnetic susceptibility and estimates of the Luttinger parameter \(K\) from fits to the Rényi entropy, see text.
        The solid red line indicates the short-range analytic result.
        Data for \(\alpha^{-1} \leq 0.3\), color coded from dark blue to bright yellow while data for \(\alpha^{-1} > 0.3\) is color coded from bright yellow to dark brown.
        The uncertainty on each data point is marked by vertical capped error bars.
    }
\end{figure}

We finally consider the behavior of the speed of sound \(u\) in the LL regime.
To find it, we rely on the LL formula for the magnetic susceptibility
\(\chi = \evaluated{\pdv*{M}{h}}_{h=0} =  {K}/(u \pi)\), where \(h\) is the magnetic field amplitude and \(M = 2\sum_n \expval{S^z_n}\) is the total magnetization.
In the MPS simulations,
we add the magnetic coupling term \(-h \sum_n S^z_n\)  to Hamiltonian~\eqref{eq:H-LRXXZ} and compute \(M\) for various magnetic field amplitudes in the range \(0.01 < h < 0.1\).
The magnetic susceptibility \(\chi\) is then found from a linear fit to the MPS data
and the speed of sound as \(u = K^\mathcal{S}/(\pi \chi)\), where \(K^\mathcal{S}\) is the estimate of the Luttinger parameter found from the entanglement entropy as discussed above.
The results (colored disks), 
together with the Bethe ansatz prediction
\(u(\Delta) = \pi \sqrt{1-\Delta^2}/(2\arccos(-\Delta))\)~\cite{takahashi_minoru_thermodynamics_1999,giamarchi_quantum_2003,franchini_introduction_2017} are shown in Fig.~\ref{fig:luttinger-u}.
In striking contrast with the Luttinger parameter \(K\), the speed of sound \(u\) does not follow the rescaling of \cref{eq:rescaling}. It rather increases in value upon increasing the long-range parameter \(\alpha^{-1}\).
Quantitatively, the increase goes from a few percent for \(\alpha^{-1}=0\) to 
\(\simeq 30\%\) for \(\alpha^{-1} \simeq 0.2\)
marking a small yet appreciable mismatch with the analytic short-range prediction, see Fig.~\ref{fig:luttinger-u} data colored from dark blue to bright green.
In contrast, for longer range interactions, \(\alpha^{-1} \gtrsim 0.2\), we find a dramatic increase of \(u\) with respect to the short-range value, see Fig.~\ref{fig:luttinger-u} data colored coding from bright yellow to black.

Note that the fact that \(u\) does not fulfill the same rescaling as \(K\) does not call into question the validity of the mapping to the effective short-range LL model identified above.
It, however, indicates that the long-range XXZ Hamiltonian in the LL phase cannot be mapped into its short-range equivalent solely by the rescaling~\eqref{eq:rescaling}.
In fact, it is necessary to add a rescaling in energy, determined by the results of Fig.~\ref{fig:luttinger-u}.
Since the latter only affects the energy scale, it does not affect the entanglement Hamiltonian.

\section{\label{sec-conclusion}Conclusion}
In this paper, we have shown that the entanglement properties, and more precisely the entanglement spectrum, are instrumental in determining both first-order and infinite-order phase transitions in a long-range quantum spin model.
Specifically, we have shown that the entanglement spectrum contains sufficient information to fully determine the quantum phase diagram of the LRXXZ model,
and locate the corresponding phase transitions, vs the anisotropy and long-range parameters.
In contrast, geometrical entanglement signals the AFM--SRXY and SRXY--FM transitions, reminiscent of the short-range XXZ model, but shows a smooth behavior at the onset of genuine long-range effects, namely across the SRXY-LRXY transition.
We have found that, within the XY phase, the entanglement spectrum exhibits a remarkable self-similarity, which allows us to map the long-range model onto its short-range counterpart. The latter can be exploited in combination with RG theory to locate the AFM--SRXY and SRXY--LRXY phase transitions from the breakdown of LL theory.
The AFM, SRXY, and FM phases hence obtained are reminiscent of the short-range XXZ model, while the LRXY phase is characterized by emerging long-range effects and continuous symmetry breaking. The obtained phase diagram is in good agreement with our numerical calculations using tensor-network approaches, as well as with previous results.

We have further shown that the self-similarity identified in the entanglement properties extends to both the geometrical entanglement and the Luttinger parameter in the SRXY phase. In contrast, the speed of sound, which defines the energy scale of LL theory exhibits a different rescaling with the long-range parameter. Finally, we have checked the validity of LL theory by comparing estimates of the Luttinger parameter from various Rényi entropies and correlation functions, which all agree within the SRXY phase.

These results call for further studies of the entanglement properties of long-range quantum systems. 
A particularly important question would be to understand the origin of the self-similar rescaling found here from a microscopic point of view including effects beyond perturbation theory, and extend it to other quantum models as well as thermal equilibrium states.
We expect that the approach we use here can be straightforwardly extended to other short-range and long-range models. In this respect, we stress that the analysis of the entanglement spectrum is self-contained and does not rely on any previous knowledge of the phase diagram. Degeneracy lifts are the primary signals of phase transitions and appear to be robust against finite-size effects. Other features, such as the Schmidt gap, may be more sensitive to finite-size effects but can be excluded using proper finite-scaling analysis.
The approach developed here may also constitute a useful tool in studying out-of-equilibrium dynamics of many-body quantum systems with long-range interactions, which attracts significant attention~\cite{schachenmayerEntanglementGrowthQuench2013, haukeSpreadCorrelationsLongRange2013, eisertBreakdownQuasilocalityLongRange2013,  richermeNonlocalPropagationCorrelations2014, jurcevicQuasiparticleEngineeringEntanglement2014, foss-feigNearlyLinearLight2015,cevolani_protected_2015, cevolaniSpreadingCorrelationsExactly2016, buyskikhEntanglementGrowthCorrelation2016, cevolaniUniversalScalingLaws2018, villa_unraveling_2019, despresTwofoldCorrelationSpreading2019, despresTwofoldCorrelationSpreading2019, elseImprovedLiebRobinsonBound2020, schneiderSpreadingCorrelationsEntanglement2021}.

\begin{acknowledgments}
    We thank Grégoire Misguish and Luca Tagliacozzo for stimulating discussions.
    Numerical calculations were performed using HPC resources from CPHT
    and HPC/AI resources from GENCI-CINES (Grant No.\ 2020-A0090510300) and GENCI-TGCC (Grant No.\ 2021-A0110510300).
    The DMRG calculations were performed using the ITensor library for the Julia programming language~\cite{fishmanITensorSoftwareLibrary2020}.
\end{acknowledgments}

\appendix

\section{FM phase ground state analysis}\label{app:fm-groun-state}
In this appendix, we provide a simple argument showing that the ground state of the LRXXZ model in the FM phase (\(\Delta \geq 1\))
is a trivial, fully polarized, product state. Proofs at the spin-wave level are discussed in previous papers, see for instance Refs.~\cite{frerot_entanglement_2017,maghrebi_continuous_2017}.
Here we write the Hamiltonian~\eqref{eq:H-LRXXZ} as
\begin{align} \label{eq:H-LRFM}
    H_\mathrm{LRXXZ} &=  \sum_{n\neq m} \frac{-{J}/{2}}{\abs{n-m}^\alpha}\left( \vec{T}_{nm}^2 + (\Delta -1) (T^z_{n,m})^2 \right) + \mathrm{const} \,, 
\end{align}
where
we have introduced the two-site spin operator \(\vec{T}_{nm} = \vec{S}_n + \vec{S}_m\),
\({T}^z_{nm} = {S}^z_n + {S}^z_m\),
and we have used the identities \({\vec{S}_n}^2 = S(S+1) = 3/4\) and \(({S}^z_n)^2 = 1\).
Consider first the minimization of the energy of each two-site term independently, corresponding to the maximization of the expectation value,
\begin{align}\label{eq:pair}
    \expval{ \vec{T}_{nm}^2 + (\Delta -1) (T^z_{nm})^2}{\psi_{n,m}} \longrightarrow \max_{\psi_{n,m}}.
\end{align}
The operator \(\vec{T}_{nm}\) represents a spin \(0\) or spin \(1\), and the quantity \(\expval*{\vec{T}_{nm}^2}\) is maximized for spin \(1\).
The quantity \(\expval*{ (T_{nm}^z)^2}\) is also maximized for spin \(1\) configurations. Either of the product states \(\ket*{T=1, T_z=+1}_{n,m} = \ket*{\uparrow}_n \otimes \ket*{\uparrow}_m\) or \(\ket*{T=1, T_z=-1}_{n,m} =  \ket*{\downarrow}_n \otimes \ket*{\downarrow}_m\) represent spin 1 for \(\vec{T}_{nm}\) and maximize the pair term of \cref{eq:pair} for \(\Delta \geq 1\). It follows that either of the fully polarized states
\(\ket*{\psi} = {\bigotimes}_n \ket*{\uparrow}_n\) or \(\ket*{\psi} = {\bigotimes}_n \ket*{\downarrow}_n\) jointly maximizes all the pair terms and consequently minimize \(H_\mathrm{LRXXZ}\).
Note that for \(\Delta <1\), each pair is optimized by the antiferromagnetic state
\(\ket*{T=1, T_z=0}_{n,m} =  \big(\ket*{\uparrow}_n \otimes \ket*{\downarrow}_m + \ket*{\downarrow}_n \otimes \ket*{\uparrow}_m\big)/\sqrt{2}\). This yields frustration when including all two-site terms and pairwise optimization for the entire Hamiltonian breaks down.

\section{Finite-size scaling}\label{app:finite-size}
\begin{figure}
    \centering
    \includegraphics[width=0.9\linewidth]{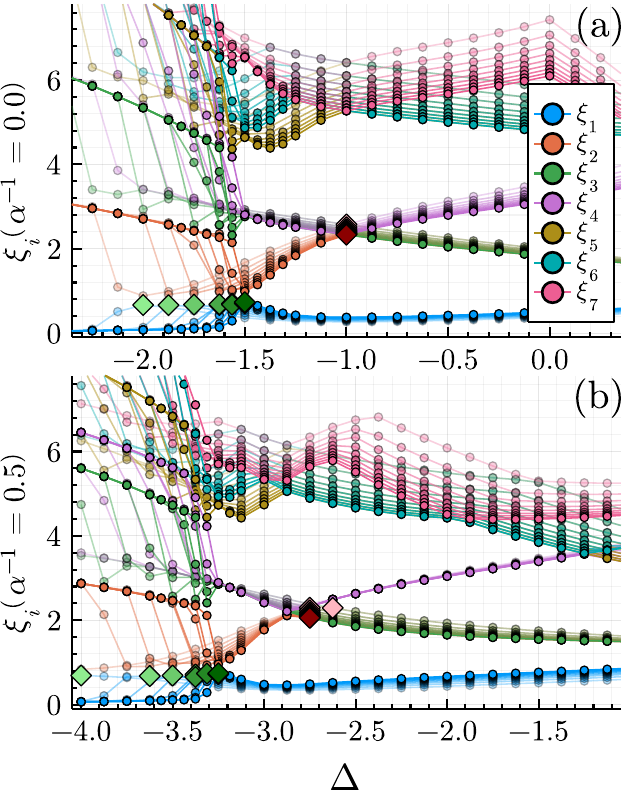}
    \caption{Finite-size scaling of the first seven entanglement energies \(\left\{ \xi_j \right\}\) for
    (a)~\(\alpha^{-1}=0\) and
    (b)~\(\alpha^{-1}=0.5\). 
    The transparency of each line decreases with the system size from \(N=60\) (\(70\%\) transparency) to \(N=220\) (\(0\%\) transparency). The size step is \(\Delta N = 20\). The degeneracy lift is marked by the red diamonds for each system size (light red \(N=60\), dark red \(N=220\)), and the minimum of the Schmidt gap (\(\xi_2 - \xi_1\)) is marked by green diamonds for the corresponding system size (light green \(N=60\), dark green \(N=220\)).
    }
    \label{fig:ES-finite-size}
\end{figure}

In this appendix, we examine the finite-size scaling of the bare observables on the MPS ground state.
To this end, we showcase the explicit finite-size scaling of the short-range (\(\alpha^{-1}=0\)) and a long-range (\(\alpha^{-1} = 0.5\)) case for the geometric entanglement as well as the entanglement spectrum.
{The same qualitative behavior and convergence are found for all other \(\alpha^{-1}\) considered in the main text too}.
Here, we inspect system sizes \(N = [60,80,\ldots,220]\).
\begin{figure}[h!]
    \centering
    \includegraphics[width=0.9\linewidth]{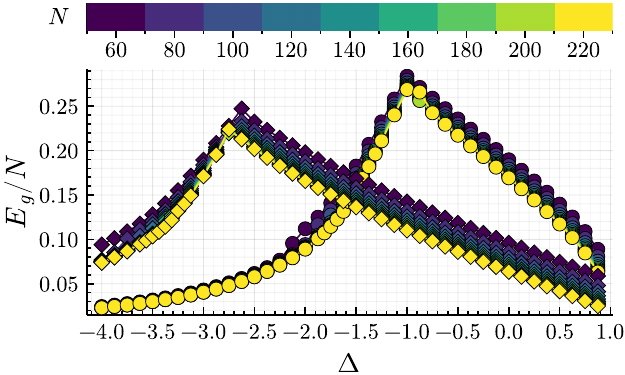}
    \caption{\label{fig:GE-finite-size}
        Geometric entanglement density \(E_g/N\) over the anisotropic parameter \(\Delta\) for \(\alpha^{-1}=0\) (colorized circles) and \(\alpha^{-1}=0.5\) (colorized diamonds)
        whereas the color corresponds to system size indicated in color bar above.
    }
\end{figure}
As illustrated in \Cref{fig:GE-finite-size}, the GE is well converged in system size for \(N \geq 160\) for all considered values of \(\alpha^{-1}\).
The finite-size scaling of the entanglement spectrum is showcased in \cref{fig:ES-finite-size}.
Although we observe the ES being moderately well converged for the largest system sizes and large entanglement energies, the degeneracy lift signaling the AFM-XY phase transition is very well converged as are the low entanglement energy lines, cf.\ red diamonds in \cref{fig:ES-finite-size}.
The Schmidt gap, defined as the difference \(\xi_2 - \xi_1\), has a local minimum which is marked by the green diamonds in \cref{fig:ES-finite-size}, which directly corresponds to a local maximum in the entanglement entropy. Furthermore, this local minimum coincides with a second lift of the apparent degeneracy of \(\xi_3 = \xi_4\).
We refer to this region between the local minimum of the Schmidt gap and the lift of the degeneracy \(\xi_2 = \xi_3\) as the crossover regime in the main text.
\begin{figure}
    \centering
    \includegraphics[width=0.9\linewidth]{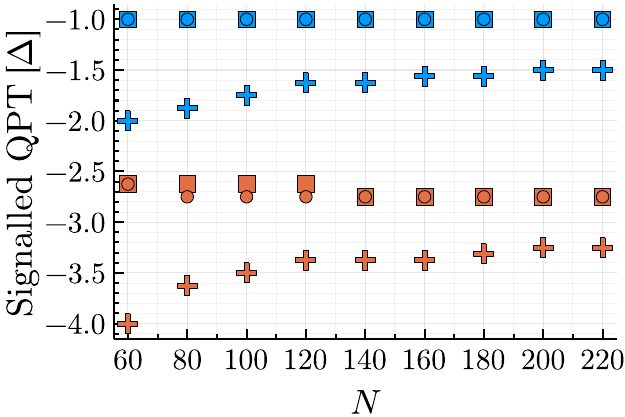}
    \caption{Position of AFM-XY quantum phase transition (QPT) in values of the anisotropy coupling \(\Delta\) vs system sizes \(N\) as found from the degeneracy lift in the ES (circles), from the maximum of the GE (squares), and from the local minimum of the Schmidt gap (\(\xi_2 - \xi_1\), crosses).
    Blue and orange markers correspond to \(\alpha^{-1}=0\) and \(\alpha^{-1}=0.5\), respectively.}
    \label{fig:QPT-signals-finite-size}
\end{figure}
\Cref{fig:QPT-signals-finite-size} shows the finite-size convergence of these independent signals and the size of the crossover regime. In the short-range case, the two signals locate the QPT at the same position for all system sizes considered (blue data in \cref{fig:QPT-signals-finite-size}).
Contrarily, we find the degeneracy lift being converged much earlier in system size compared to the local maximum of \(E_g/N\) when considering longer ranged interactions, cf.\ orange data in \cref{fig:QPT-signals-finite-size}.
Furthermore, we observe a very slow convergence of the crossover regime in system size, cf.\ crosses in \cref{fig:QPT-signals-finite-size}, consistently with previous studies of slow entanglement entropy convergence of the (short-range) XXZ model~\cite{wang_berezinskii-kosterlitz-thouless_2010,li_ground_2019}.

\section{Optimization of self-similar rescaling parameters}\label{app:optim-parameters}

To obtain the optimal self-similar rescaling parameters for the ES, we proceed as follows. We first consider each spectral line as a function of the anisotropy parameter, \(\xi_j = \xi_j(\Delta)\) and rescale the argument as \(\Delta \rightarrow \tilde{\Delta}(\gamma, \nu)\) following \cref{eq:rescaling}.
The rescaling parameters \(\gamma(\alpha)\) and \(\nu(\alpha)\) depend on the anisotropy parameter \(\alpha\), and we have the fixed point \(\tilde{\Delta}(\alpha = \infty) = \Delta\).
We compute the ES of the MPS ground state for a discrete set of values of the anisotropy parameter, \(\Delta_d = -4, -3.875, \ldots 1.5\)
and interpolate linearly each spectral line.
The scaling parameters \(\gamma(\alpha)\) and \(\nu(\alpha)\) are then fitted so as to minimize the quadratic weight function over a few (here \(4\)) low-lying spectral lines,
\begin{align}
    W(\gamma, \nu) &= \sum_{1 \leq j \leq 4} \int_{\tilde{\Delta}_i}^{\tilde{\Delta}_f} \dd{\tilde{\Delta}} \left[ \xi_j\big(\tilde{\Delta}(\gamma,\nu),\alpha\big)  - \xi_j(\Delta,\alpha=\infty) \right]^2 \,,
\end{align}
where
we vary the integral boundaries \(\tilde{\Delta}_i\), \(\tilde{\Delta}_f\) over a set of four different pairs of lower and upper integral boundaries \((\tilde{\Delta}_i,\ \tilde{\Delta}_f) = [(-1.5, 0.0), (-1.25, -0.5), (-1.25, 0.5), (-2, 0.5)]\) such as to optimize the rescaling over a broad region including the AFM--XY transition, at \(\tilde{\Delta} = -1\).
We then infer the value of the rescaling parameters \(\gamma\) and \(\nu\) from the average over the four samples of different integration intervals and estimate its error from the standard deviation.
The result is shown in \cref{fig:rescaling-params}  
The numerical optimization is performed with the Julia package Optim.jl~\cite{mogensen_optim_2018}.

\section{RG flow analysis of the LRXXZ}\label{app:RG-flow}
In this appendix, we present a derivation of the RG flow equations of operators \(V_\mathrm{ ZZ}\) and \(V_\mathrm{ LRXY}\) from \cref{eq:V-ZZ,eq:V-LRXY}.
Following textbook approaches for the bosonization of the XXZ model~\cite{giamarchi_quantum_2003,sachdev_quantum_2011,franchini_introduction_2017,shankar_quantum_2017}, we start with the LRXXZ Hamiltonian~\eqref{eq:H-LRXXZ} at the free fermion point \(\Delta=0\), \(\alpha=\infty\).
It is expressed using the Jordan--Wigner transform as
    \(H_{0} = -\frac{J}{2} \sum_{n} c^\dagger_n c_{n+1} + \mathrm{h.c.}
    = -{J} \int_{-\frac{\pi}{a}}^{\frac{\pi}{a}} \frac{\dd{k}}{2\pi} \cos(ka) \, \tilde{c}^\dagger_k \tilde{c}_{k}\)
where \(c_n\) is the fermionic annihilation operator on site \(n\), \(\tilde{c}_k\) the Fourier transform thereof, \(\acomm*{c_n}{c^\dagger_m} = \delta_{n,m}\), and we identify the dispersion relation \(\omega(k) = -J \cos(ka)\).
The free fermions have two Fermi points at  \(\pm k_F\) with \(k_F= \pi/(2a)\).
The low energy modes up to a cutoff \(1/a_{\mathrm{c}}\) around either Fermi point are then approximated by a linear dispersion relation \(\eval*{\omega(k)}_{\pm k_F} \simeq \pm J a (k \mp k_F)\).
It gives rise to the Fourier mode decomposition of a fermionic quantum field with two species,
    \(\psi_\pm(x_n) = \lim_{a \rightarrow 0} \frac{1}{\sqrt{a}} c_n \exp(\pm \i k_F x_n)\)\,,
with \(x_n = a n\). We can then formulate \(H_0\) to read as,
   \( H_0 = \frac{u}{2} \int \dd{x} \left[ K \left(\Pi(x)\right)^2 + \frac{1}{K} \left(\partial_x \phi(x)\right)^2 \right]\),
by means of the bosonization formula
    \(\psi_{\pm }(x) \equiv \frac{1}{\sqrt{2 \pi a_{\mathrm{c}}}} \exp[ \mp \i  \sqrt{4\pi} \phi_\pm(x)] \),
and the definitions \(\phi(x) = \phi_+(x) + \phi_-(x)\), \(\theta(x) = \phi_+(x) - \phi_-(x)\)
and \(\Pi(x) = \partial_x \theta(x)\) the canonical conjugate of \(\phi\), \(\comm*{\Pi(x)}{\phi(y)}= \i \delta(x-y)\).
In the XY case, \(\Delta=0\), we have
\(K=1\) and \(u=aJ\), which is nothing but the Fermi velocity.
The action associated to \(H_0\) reads as
\begin{align}
    S_0[\phi] &= \frac{1}{2K} \int \dd[2]{\mathbf{r}} \left( \partial_\mu \phi(\mathbf{x}) \right)^2 \\
    &= \frac{1}{2K} \int \frac{\dd[2]{\mathbf{p}}}{(2\pi)^2} \left( p^\mu \phi(\mathbf{p}) \right)^2 \\
    &\eqqcolon \frac{1}{2} \int \frac{\dd[2]{\mathbf{p}}}{(2\pi)^2}  \phi(p) D_\phi^{-1}(p) \phi(p)
\end{align}
with \(\mathbf{r} = r^\mu = (u \cdot t, x)\) the spacetime vector (with Euclidean norm), \(\mathbf{p} = p^\mu = (\omega /u, k)\) the spacetime Fourier vector, and \(D_\phi(p) = K/p^2\) the (free) Feynman propagator for field \(\phi\).
Subsequently, we relate the microscopic spin degrees of freedom on the lattice to their bosonic counterpart in the continuum as~\cite{giamarchi_quantum_2003,franchini_introduction_2017,shankar_quantum_2017},
\begin{align}
    S^+(x) &= \frac{S^\pm_n}{\sqrt{a}}  = \frac{(-1)^{x/a}}{\sqrt{2\pi a_{\mathrm{c}}}}  \exp(\pm \i  \sqrt{\pi} \theta(x))   \,, \\
    S^z(x) &= \frac{S^z_n}{a_{\mathrm{c}}}  = \frac{1}{\sqrt{\pi}} \partial_x \phi(x) - \frac{(-1)^{x/a_{\mathrm{c}}}}{\pi a_{\mathrm{c}}} \sin(\sqrt{4 \pi} \phi(x)) \,,
\end{align}
where the Jordan--Wigner phase factor \(\Phi(x)\) is taken to be the Hermitian version \(\Phi(x) =  \frac{1}{2} \exp(-\i \pi \sum_{y<x} c^\dagger_y c_y ) + \mathrm{h.c.} = \cos(\phi(x) - k_F x)\).
We now perturbatively include interaction terms when \(\Delta \neq 0\) and \(\alpha < \infty\) based on the free field definitions. To this end, we take the standard textbook result form the bosonization of \(-J \Delta \sum_n S^z_n S^z_{n+1}\)~\cite{{giamarchi_quantum_2003,franchini_introduction_2017,shankar_quantum_2017}}
\begin{align}
   V_\mathrm{ ZZ} &= - J \Delta \int \dd{x} \left[ \frac{4}{\pi} \left( \partial_x \phi(x)\right)^2
   + \frac{1}{(\pi a_{\mathrm{c}})^2} \cos(\sqrt{16\pi} \phi(x)) \right] \,,
\end{align}
and include the long-range interaction in XY-direction \(\sum \sum_{n,m,n\neq m}  \left( S^+_n S^-_m +\mathrm{h.c.} \right)/\left({2\abs*{n-m}^\alpha}\right) \) as 
\begin{align}
    V_\mathrm{ LRXY}  &= - \frac{J}{2\pi a_{\mathrm{c}}} \int_{\abs*{x-y}\gg a}  {\cos\left(\sqrt{\pi} \big[\theta(x)-\theta(y)\big] \right)} \times \nonumber \\
    & \qquad \times \frac{(-1)^{(x-y)/a}}{\abs*{x-y}^\alpha}  \dd{x} \dd{y} \,.
\end{align}
Note that the long-range interaction term \(S^z_n S^z_m/\abs*{n-m}^\alpha\) is highly irrelevant for \(\alpha^{-1}<1\)~\cite{inoue_conformal_2006},
and we omit it in our RG analysis.
Included in the short-range \(S^z S^z\) interaction is an additional term quadratic in bosonic fields.
This term originates in the point-splitting procedure which takes into account that square and higher terms of the fields in coordinate space are not defined and have to be regularized by the microscopic lattice~\cite{franchini_introduction_2017}.
This term perturbatively renormalizes the free bosonic Hamiltonian according to \(u K = v_F\), and \({u}/{K} =  J \left(   1+ {4\Delta}/{\pi}\right)\) for the parameters in Hamiltonian~\(H_0\). 

We subsequently consider the microscopic generating functional \(Z = \exp(-S[\phi,\theta])\) with action \(S[\phi,\theta] = S_0[\phi] - V_\mathrm{ ZZ}[\phi] - V_\mathrm{ LRXY}[\theta]\). 
We split the spacetime Fourier modes of both fields in slow and fast moving ones according to,
\begin{align}
    \phi(k) &= \begin{cases}
      \phi_s(k) &\quad \text{if} \quad  0 \leq k \leq \Lambda(1-\dd{\ell}) \\
      \phi_f(k) &\quad \text{if} \quad \Lambda(1-\dd{\ell})< k \leq \Lambda
    \end{cases}  \,, \\
    \phi(k) &= \phi_s(k) + \phi_f(k) \,,
  \end{align}
with \(k\) being the norm of the Fourier spacetime vector, and \(\dd{\ell}\) the width of the spacetime momentum shell being integrated out. The analogous splitting applies to the field \(\theta(k) = \theta_s(k) + \theta_f(k)\).
Including interaction terms, we integrate out the fast moving modes in the momentum shell \(\dd{\ell}\) and inspect how the coupling constants change under such an RG step, which yields,
\begin{widetext}    
\begin{align}
  Z &= \int \int  \DD{\phi_s} \DD{\phi_f} \DD{\theta_s} \DD{\theta_f} \exp( -\int \left[ \frac{1}{2}(\partial_\mu \phi_s)^2 + \frac{1}{2}(\partial_\mu \phi_f)^2 \right] \dd{t} \dd{x} ) \exp(- \frac{g_\mathrm{ ZZ} \Lambda^2}{2} \int \cos(\sqrt{16\pi}\left(\phi_s + \phi_f \right)) \dd{\tau} \dd{x}) \nonumber \\
  &\quad \times \exp(-\frac{g_{LR} \Lambda}{2} \int \frac{(-1)^{(x-y)/a}}{\abs*{x-y}^\alpha} \cos(\sqrt{{\pi}} \left[\theta_s(t,x) - \theta_s(t,y) + \theta_f(t,x) - \theta_f(t,y)\right]) \dd{t} \dd{x} \dd{y}) \,,
\end{align}
where we identified \(\Lambda = 1/(\pi a_{\mathrm{c}})\).
Note that under the path integral the fields are only \(\mathbb{C}\)--numbers and thus commute.
Next we expand \(\cos(a+b) = \cos(a)\, \cos(b) - \sin(a)\, \sin(b)\) and ignore the terms proportional to \(\sin(\phi_f)\) or \(\sin(\theta_f)\)
because they average to zero over the even path integral measure.
Next, we recognize the expectation value with respect to the ground state of \(\phi_f\) defined as
\begin{align}
\expval{A[\phi_f]} = \int \DD{\phi_f} A[\phi_f] \exp(-S_0^f[\phi_f]) = \int \DD{\phi_f} A[\phi_f] \exp( -\frac{1}{(2\pi)^2} \int_{\Lambda(1-\dd{\ell})}^\Lambda  (p^\mu \phi_f)^2  \dd{\mathbf{p}} )
\end{align}
where the spacetime integral over the action only contains modes in the momentum shell of width \(\dd{\ell}\).
This yields
\begin{align}
    Z &= \int  \DD{\phi_s} \exp( -\int \frac{1}{2}(\partial_\mu \phi_s)^2 \dd{\tau} \dd{x} ) \times  {\Bigg \langle} \exp[- \frac{g_\mathrm{ ZZ} \Lambda^2}{2} \int \cos(\sqrt{16\pi} \phi_s ) \cos(\sqrt{16\pi}
    \phi_f) \dd{\tau} \dd{x}] \nonumber \\
    &\quad \times \exp[-\frac{g_{LR} \Lambda}{2} \int \frac{(-1)^{(x-y)/a}}{\abs*{x-y}^\alpha} \cos\left(\sqrt{{\pi}} \left[\theta_s(t,x) - \theta_s(t,y) \right]\right) \cos\left(\sqrt{{\pi}} \left[\theta_f(t,x) - \theta_f(t,y)\right]\right) \dd{t} \dd{x} \dd{y}]  {\Bigg \rangle}_f \,.
\end{align}
The expression above is exact and would yield the full, non-perturbative picture of a renormalization step. However, it is  unfeasible to compute \(\expval{\e^A}\) non-perturbatively.
Hence, we introduce an approximation in the form of a first-order cumulant expansion, \(\expval{\e^A} \approx \e^{\expval{A}}\), thereby ignoring higher order cross-terms of the interaction operators,
\begin{align}
  Z &\approx \int  \DD{\phi_s} \exp( -\int \frac{1}{2}(p^\mu \phi_s)^2 \frac{\dd[2]{p}}{(2\pi)^2} ) \times \exp(- \frac{g_\mathrm{ ZZ} \Lambda^2}{2} \int \cos(\sqrt{16\pi} \phi_s ) \expval{\cos(\sqrt{16\pi}
  \phi_f)}_f \frac{\dd[2]{p}}{(2\pi)^2}) \nonumber \\
  &\quad \times \exp(-\frac{g_{LR} \Lambda}{2} \int \frac{(-1)^{(x-y)/a}}{\abs*{x-y}^\alpha} \cos\left(\sqrt{{\pi}} \left[\theta_s(t,x) - \theta_s(t,y) \right]\right) \expval{\cos(\sqrt{{\pi}} \left[\theta_f(t,x) - \theta_f(t,y)\right])}_f \dd{t} \dd{x} \dd{y})  \,.
\end{align}
To evaluate the expectation value of a trigonometric function of the fields, we use below identity~\cite{shankar_quantum_2017},
\begin{align}
  \e^A \cdot \e^B &=  \normord{ \e^{A + B}} \e^{\expval{A B} + \frac{1}{2}\expval{A^2 + B^2}} \,, \label{eq:BCH}
\end{align}
where \(\normord{A}\) is the normal ordering of \(A\).
It follows as a corollary from the Baker--Campbell--Hausdorff formula when \(\comm*{A}{B}\) commutes with \(A\) and \(B\).
Note that we have by definition \({\normord{A}}\ket{0} = 0\) implying \(\expval{ \normord{\exp(A)}} = 1\).
Using \cref{eq:BCH} and setting \(B=0\) and \(A=\i \beta X\), we find
\begin{align}
  \expval{\e^{\i \beta X}} = \e^{-\frac{1}{2} \beta^2 \expval{X^2}} \,, \quad 
  \text{implying} \quad \expval{\cos(\beta X)} = \e^{-\frac{1}{2} \beta^2 \expval{X^2}} \,.
\end{align}
We thus find
\begin{align}
  \expval{\cos(\sqrt{16 \pi} \phi_f)}_f &= \exp[-8{\pi}\expval{\phi_f^2}_f] \label{eq:expval-sine-Gordon}
\end{align}
and
\begin{align}
  \expval{\cos( \sqrt{{\pi}} \left[\theta_f(t,x) - \theta_f(t,y)\right]) }_f &= \exp[-\frac{\pi}{2}\expval{\left[\theta_f(t,x) - \theta_f(t,y)\right]^2}_f] \,. \label{eq:expval-LR}
\end{align}
\Cref{eq:expval-sine-Gordon} is readily evaluated as~\cite{shankar_quantum_2017}
\begin{align}
    \expval{\cos(\sqrt{16 \pi} \phi_f)}_f &= \exp[-8\pi \expval{\phi_f^2}_f] = \exp[ -8\pi \frac{1}{(2\pi)^2}  \int_{\Lambda(1-\dd{\ell})}^{\Lambda}  D_\phi(\mathbf{p}) \dd[2]{\mathbf{p}}]  \nonumber \\
    &= \exp[-8\pi \frac{1}{(2\pi)^2} \int_{\Lambda(1-\dd{\ell})}^{\Lambda} \frac{K}{p^2} p \dd{p}  \int_0^{2\pi} \dd{\varphi} ] = \exp[-4  K \ln( \frac{\Lambda}{\Lambda(1-\dd{\ell})})] = 1 - 4K \dd{\ell}\,.
\end{align} 
Upon rescaling with \(s = \Lambda/\Lambda'\) and \(\Lambda'=(1-\dd{\ell})\Lambda\), the spacetime integral measure reads \(\dd[2]{\mathbf{x}} = s^2 \dd[2]{\mathbf{x}'} = (1+2\dd{\ell})\dd[2]{\mathbf{x}'}\).
With this result, we conclude that one RG step yields the RG flow equation for \(g_\mathrm{ ZZ}\) up to first order perturbation theory (omitting to prime new variables),
\begin{align}
    \frac{g_\mathrm{ ZZ}\Lambda^2}{2} \int \dd[2]{\mathbf{x}} \cos(\sqrt{16\pi}\phi(\mathbf{x})) &\rightarrow  \frac{g_\mathrm{ ZZ}\Lambda^2}{2} \left(1 + \left(2-4K\right)\dd{\ell}\right)\int \dd[2]{\mathbf{x}} \cos(\sqrt{16\pi}\phi(\mathbf{x})) \,.
\end{align}
Hence, we find
\begin{align}
\dv{g_\mathrm{ ZZ}}{l} &= \left(2-4K\right) g_\mathrm{ ZZ} \,,
\end{align}
which is \cref{eq:rg-flow-zz} in the main text.

\Cref{eq:expval-LR} is similarly evaluated as the connected two-point equal-time correlation function.
With the use of the symmetry of \(S_0\) under the duality transformation \(\phi \rightarrow \theta\), \(K \rightarrow 1/K\), \(D_\phi(p) \rightarrow D_\theta(p) = K^{-1} p^{-2}\), \cref{eq:expval-LR} yields~\cite{giamarchi_quantum_2003,franchini_introduction_2017,shankar_quantum_2017},
\begin{align}
    \expval{\cos( \sqrt{{\pi}} \left[\theta_f(t,x) - \theta_f(t,y)\right]) }_f &= \exp[-\frac{\pi}{2}\expval{\left[\theta_f(t,x) - \theta_f(t,y)\right]^2}_f] \\
    &= \exp[ -\frac{\pi}{2} \int_{\abs{x - y} \gg a} \expval{\theta_f(\mathbf{p})\theta_f(\mathbf{q})}_f \left(\e^{\i \mathbf{p} \mathbf{x}} - \e^{\i \mathbf{p}\mathbf{y}}\right)\left(\e^{\i \mathbf{q} \mathbf{x}} - \e^{\i \mathbf{q}\mathbf{y}}\right)  \dd[2]{\mathbf{p}} \dd[2]{\mathbf{q}} ]\\
    &= \exp[ -\frac{\pi}{2} \frac{1}{(2\pi)^2}  \int_{\Lambda(1-\dd{\ell})}^{\Lambda} D_\theta(p) \delta(\mathbf{p}+\mathbf{q})  \left(\e^{\i \mathbf{p} \mathbf{x}} - \e^{\i \mathbf{p}\mathbf{y}}\right)\left(\e^{\i \mathbf{q} \mathbf{x}} - \e^{\i \mathbf{q}\mathbf{y}}\right) \dd[2]{\mathbf{p}} \dd[2]{\mathbf{q}}  ]  \\
    &= \exp[ -\frac{\pi}{2} \frac{1}{2\pi}  \int_{\Lambda(1-\dd{\ell})}^{\Lambda} 2 D_\theta(p)  \left[1 - \cos(p \norm{\mathbf{x}-\mathbf{y}})\right] p \dd{p}   ]  \\
    &= \exp[-\frac{1}{2K} \ln( \frac{\Lambda}{\Lambda(1-\dd{\ell})})] = 1 - \frac{\dd{\ell}}{2K}\,.
\end{align} 
Here above we ignored the integral over the cosine since its frequency oscillations are large \(\norm{\mathbf{x}-\mathbf{y}} \gg a\) compared to the modes considered for \(p \simeq \Lambda = 1/a\), and it thus averages out under the integral.  
In the case of \(g_\mathrm{LRXY}\), the spacetime integral measure transforms as \(\dd{t}\dd{y}\dd{x} = s^3 \dd{t'}\dd{y'}\dd{x'} = (1+3\dd{\ell})\dd{t'}\dd{y'}\dd{x'}\) while the long-range interaction potential scales as \(\abs*{x-y}^{-\alpha} = s^{-\alpha}\abs*{x'-y'}^{-\alpha} = (1-\alpha\dd{\ell})\abs*{x'-y'}^{-\alpha}\).
Therefore, the RG step and RG flow equation for the long-range XY operator yield (omitting to prime new variables),
\begin{align}
     -\frac{g_{LR} \Lambda}{2} \int \frac{\cos(\sqrt{{\pi}} \left[\theta(t,x) - \theta(t,y)\right])}{\abs*{x-y}^\alpha} &\rightarrow -\frac{g_{LR} \Lambda}{2} \left( 1+ \left(3-\alpha -\frac{1}{2 K}\right)\dd{\ell}\right)\int \frac{\cos(\sqrt{{\pi}} \left[\theta(t,x) - \theta(t,y)\right])}{\abs*{x-y}^\alpha} \,,
\end{align}
and 
\begin{align}
\dv{g_\mathrm{ LRXY}}{l} &= \left(3-\alpha -\frac{1}{2 K}\right) g_\mathrm{ LRXY} \,,
\end{align}
which is \cref{eq:rg-flow-lr} in the main text.
\end{widetext}

\bibliography{Bibliography.bib}

\end{document}